\documentclass[12pt]{article}

\usepackage{graphicx}
\begin{document}
\newcommand{\NP}[1]{Nucl.\ Phys.\ {\bf #1}}
\newcommand{\PL}[1]{Phys.\ Lett.\ {\bf #1}}
\newcommand{\PR}[1]{Phys.\ Rev.\ {\bf #1}}
\newcommand{\PRL}[1]{Phys.\ Rev.\ Lett.\ {\bf #1}}
\newcommand{\PREP}[1]{Phys.\ Rep.\ {\bf #1}}
\newcommand{\PTP}[1]{Prog.\ Theor.\ Phys.\ {\bf #1}}
\newcommand{\MPL}[1]{Mod.\ Phys.\ Lett.\ {\bf #1}}
\newcommand{\IJMP}[1]{Int.\ Jour.\ Mod.\ Phys.\ {\bf #1}}
\newcommand{\JHEP}[1]{JHEP\ {\bf #1}}
\begin{titlepage}
\setcounter{page}{0}
\begin{flushright}
\end{flushright}
~\\
\vspace{10mm}
\begin{center}
{\Large  Causal dynamical triangulation of a 3D tensor model }

\vspace{10mm}

{\large Hiroshi\ Kawabe\footnote{e-mail address:
kawabe@yonago-k.ac.jp}} \\
{\em
National Institute of Technology, Yonago College,\\ 
Yonago 683-8502, Japan} \\
\end{center}

\vspace{8mm}
\centerline{{\bf{Abstract}}}
\vspace{5mm}
We extend the string field theory of 2D generalized causal dynamical triangulation (GCDT) with the Ishibashi-Kawai type (IK-type) interaction formulated by the matrix model to the 3D model of the surface field theory.
Based on the loop gas model, we construct a tensor model for the discretized surface field and then apply it the stochastic quantization method.
In the double-scaling limit, the model is characterized by two scaling dimensions $D$ and $D_N$, the power indices of the minimal length as the scaling parameter.
The continuum GCDT model with the IK-type interaction is realized with a similar restriction in the $D_N$-$D$ space to the 2D model.
The distinct property in the 3D model is that the quantum effect contains the IK-type interaction only, while the ordinary splitting interaction is excluded.

\end{titlepage}
\newpage
\renewcommand{\thefootnote}{\arabic{footnote}}
\setcounter{footnote}{0}
\section{Introduction}
Matrix models provide the formulation of the dynamical triangulation (DT) model describing the random surface on which discretized loops propagate and interact.
Applying the stochastic quantization method to the matrix models, string field theories are obtained through the double-scaling limit \cite{JR,Nak}.
The interaction of a loop with the spin cluster domain wall, which is called the Ishibashi-Kawai-type (IK-type) interaction, plays an important role for the construction of noncritical string field theories from the 2D matrix model \cite{IK1,IK2}.
In the loop gas model, each string is located in a domain of a 1D discrete space point $x$ and interacts with strings in the domains of only the same point or the nearest neighboring points $x \pm 1$ \cite{Kos1,KK}.
The IK-type interaction is naturally included in the matrix model formulation of the loop gas model, in which the boundary of two contiguous $x$-domains corresponds to the domain wall \cite{Kos2,Kos3}.
However, the critical problem in DT is the enormous influence of baby strings by the excessive splitting interaction \cite{AL}.
In contrast, causal dynamical triangulation (CDT) does not suffer from the above problem since it prohibits both splitting and merging interactions by the causality.
Furthermore, CDT treats the propagation better with its time-foliation structure, in which any loop propagates with an equal pace everywhere on it.
While CDT is the string propagation model, generalized CDT (GCDT) breaks the causality gently by only allowing the splitting interaction, not the merging interaction, to bring on the quantum effect \cite{ALWZ1}.
The string field theory based on GCDT is realized by the merging coupling constant zero limit of a matrix model \cite{ALWWZ1,ALWWZ2,ALWWZ3}.
The stochastic time plays the role of time, or the geodesic distance on the world sheet \cite{ALWZ2}.
GCDT is compatible with the IK-type interaction, a model of which is also constructed accordingly by a matrix model \cite{FSW}. 
In our previous work, we proposed a different class of matrix model formulation for GCDT containing the IK-type interaction, based on the loop gas model \cite{Kaw1}.
In this model, the role of the one-step discrete time $t$ is played by the 1D discrete space $x$ in the original loop gas model, instead of the stochastic time.
We discussed the scaling indices in the double-scaling limit, and they are further restricted in the open-closed string field theories, where the open string mediates the interaction of the closed string with the D-branes \cite{Kaw2}.

In the extension to the 3D model, the closed surface in each time is discretized with triangles or squares, while its propagator is composed of simplices such as tetrahedra and pyramids \cite{AJL1}.
The numerical analysis of the 3D CDT model clarifies that the stability of the propagation depends on the coupling constants of the models \cite{AJL2,AJLV,AJL3}.
On the other hand, tensor models are the natural formulation of the 3D model extended from the matrix model of 2D DT \cite{Amb,BGR}.

The question of great interest is whether the properties of GCDT are inherited in higher-dimensional models.
In this paper, we formulate a tensor model that describes the 3D GCDT model with the IK-type interaction.
The closed surface in every discrete time is discretized with squares, assigned by fundamental tensors.
In the same way as 2D CDT, any propagator is formed by accumulating shells of one-step propagators with the time-foliation structure.
Each one-step propagator is discretized with simplices, or pyramids and tetrahedra, corresponding to the basic invariant products of tensors.
In Sect.2, the tensor model is constructed by one type of space-like tensor and two types of time-like tensor.
Integrating out the time-like tensors, we obtain the effective action of the space-like tensor products, corresponding to the fields of the 2D discrete closed space, in the same form as the original loop gas model.
Because of the lack of knowledge about the one-step propagator, we include it in an effective field and investigate the scaling nature, with the details of the coefficients left unknown.
In Sect.3, applying the stochastic quantization method to the fundamental tensor, we describe the deformation of the constructive fields as quantum effects, not time evolution, with the Fokker- Planck (FP) Hamiltonian.
The IK-type interaction certainly emerges in a similar way to the 2D GCDT model.
In Sect.4, we take the double-scaling limit based on the existence of the IK-type interaction and discuss the possibility of the field theory in 3D GCDT.
The last section is devoted to the conclusions and discussions.
In the appendix, we ascertain the topological identity of the numbers of discretized simplices, and the internal links in a one-step propagator and its genus, the relation between which we use in the paper to construct the model.

\section{CDT tensor model}
With the time foliation structure of CDT, in the 3D model, any propagator of a closed surface is constructed by piling up unit time shells of one-step propagators with the same thickness everywhere.
While the closed surface field of a 2D space universe is discretized with connected squares, each shell of the 3D one-step propagator is discretized with pyramids and tetrahedra.
In the one-step time, a site may transfer to any number of connected squares through pyramids as a sweeping trajectory and vice versa.
A link, connecting two sites on the surface, transfers to any number of connected links orthogonal to the original one through sweeping tetrahedra, whose number becomes zero when two pyramids related to the link are attached with a common triangle.
Inside a piled propagator, every square belongs to two pyramids in common, or a down-pyramid in the past side and an up-pyramid in the future side.
The tetrahedron, composed of two up-triangles and two down-triangles, is necessary to mediate the connection between the up-pyramid and the down-pyramid. 
We assume that the length of any link, or an edge of these simplices, is equally the minimal length.

\begin{figure}[t]
\begin{center}
\includegraphics [width=100mm, height=50mm]{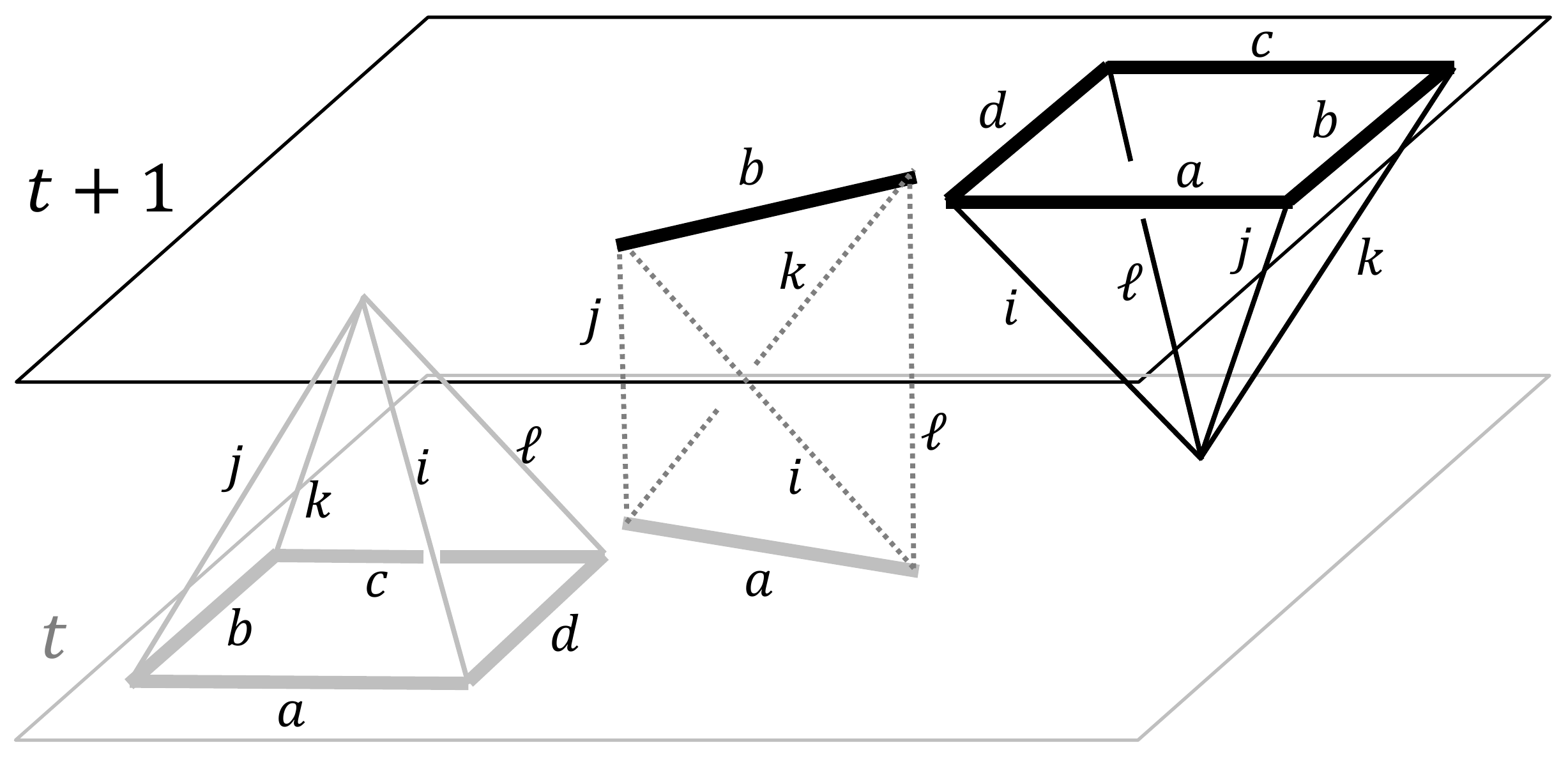}
\caption{The up-pyramid on the left-hand side: $(A^*_t)_{dcba}(B_t)_{aij}(B_t)_{bjk}(B_t)_{ck\ell}(B_t)_{d\ell i}$; the tetrahedron at the center: $(B_t)_{aij}(B_t)_{ak\ell}(C_{t+1})_{bjk}(C_{t+1})_{b\ell i}$; and the down-pyramid on the right-hand side: $(A_{t+1})_{abcd}(C_{t+1})_{aij}(C_{t+1})_{bjk}(C_{t+1})_{ck\ell}(C_{t+1})_{d\ell i}$ in the space between the time slices of $t$ and $t+1$.}
\label{fig:simplex}
\end{center}
\end{figure}

\begin{figure}[t]
\begin{center}
\includegraphics [width=100mm, height=45mm]{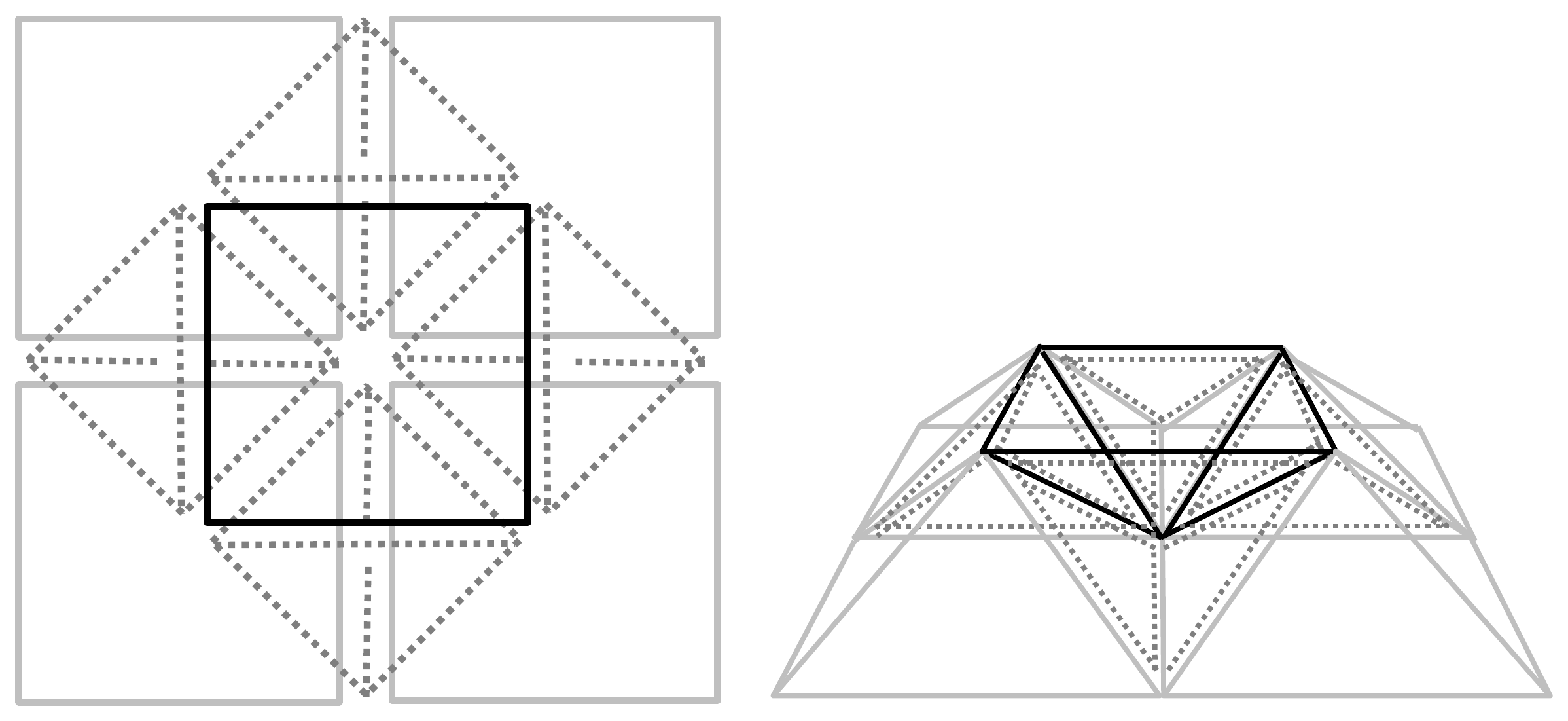}
\caption{A piece of the propagator, where four squares propagate to one square: for the connection of four up-pyramids (gray lines) and one down-pyramid (black lines), we need four tetrahedra (broken lines). The left-hand figure shows the simplices seen from the upside.}
\label{fig:foliation}
\end{center} 
\end{figure}

Here, we define a rank-four tensor, $(A_t)_{abcd}$, as the expression of a square orientable on the time slice of $t$ with the indices $a,b,c,d$ assigned to the four links in turn.
We have two types of rank-three tensor.
$(B_t)_{aij}$ corresponds to an up-triangle with one space-like link assigned $a$ lying on time $t$, while $(C_{t+1})_{aij}$ is a down-triangle with one space-like link $a$ on $t+1$, where the indices $i,j$ are assigned to time-like links.
The indices $a,b,c,d$ and $i,j$ run from 1 to $N$.
The complex tensors $A_t$, $B_t$, and $C_t$ possess the following properties:
\begin{eqnarray}
&& (A_t)_{abcd}=(A_t)_{bcda}=(A_t)_{cdab}=(A_t)_{dabc},~~~~
 (A_t^*)_{abcd} \equiv (A_t)_{dcba}, \nonumber \\
&& (B_t^*)_{aij} \equiv (B_t)_{aji}, ~~~~ (C_t^*)_{aij} \equiv (C_t)_{aji}.
\end{eqnarray} 
We start with the fundamental action of the $U(N)$ gauge-invariant form:
\begin{eqnarray}
\label{eq:action}
S[A, B, C] &=& \sum_t \left[ -Ng (A_t)_{abba} + {1 \over 2} (A_t)_{abcd} (A_t)_{dcba} - {g \over 3N} (A_t)_{abdc} (A_t)_{cdfe} (A_t)_{efba} \right. \nonumber \\
 & & + {1 \over 2} \left\{ (B_t)_{aij} (B_t)_{aji} + (C_{t+1})_{aij} (C_{t+1})_{aji} \right\} - {g \over N} (B_t)_{aij} (B_t)_{ak \ell} (C_{t+1})_{bi \ell} (C_{t+1})_{bkj} \nonumber \\
 &  &  - {g^2 \over N^2} \left\{ (A^*_t)_{dcba} (B_t)_{aij} (B_t)_{bjk} (B_t)_{ck \ell} (B_t)_{d \ell i} \right. \nonumber \\
 &  & \left. \left. ~~~~~~~+  (A_{t+1})_{abcd} (C_{t+1})_{aij} (C_{t+1})_{bjk} (C_{t+1})_{ck \ell } (C_{t+1})_{d \ell i} \right\} \right]. 
\end{eqnarray}
The invariant products, $ABBBB$ and $ACCCC$, correspond to the up-pyramids and down-pyramids, respectively, while the term $BBCC$ expresses the tetrahedra (see Fig.\ref{fig:simplex}).
The factor $g / N$ is attached to a tetrahedron, while $g^2 / N^2$ is weighted to a pyramid, regarded as a pair of tetrahedra.
The quadratic terms of $AA$, $BB$, and $CC$ work to glue the same types of face to build up a 3D space-time discretized propagator, or a foliated shell (see Fig.\ref{fig:foliation}).
The invariants of $A$ and $A^3$ cause 2D field propagation with shrinkage and expansion in the same time slice, which is not contained in CDT. 
We define the discrete surface fields composed of $n$ squares in a time slice $t$ as 
\begin{eqnarray}
\label{eq:invariants}
\phi _t (n) \equiv \left( {A_t \over N} \right) ^n , ~~~~\phi _t^{(h)} (n) \equiv {1 \over N^h} \left( {A_t \over N} \right) ^n ,
\end{eqnarray}
corresponding to the sphere, $\phi _t (n) = \phi _t^{(0)} (n)$, and the closed surface with $h$ handles, respectively.
In the above expression, all indices of the tensors are contracted in any way to make them invariants.
Integrating a product of these fields weighted by $e^{-S}$ over all kinds of tensor, we obtain any possible configuration of the propagator with these surfaces contained as its boundaries.
The partition function is
\begin{eqnarray}
\label{eq:partition}
Z = \int {\cal D} A {\cal D} B {\cal D} C e^{-S[A,B,C] } = \int {\cal D} A e^{-S_{\rm eff}[A]} ,
\end{eqnarray}
where, by integrating out only the $B$ and $C$ tensors, we may obtain the effective action $S_{\rm eff}$.
We reason out the expression of $S_{\rm eff}$ as
\begin{eqnarray}
\label{eq:effective}
S_{\rm eff} & = & S_0 + S_1 ,  \nonumber \\
S_0 & = & \sum_t \left[ {1 \over 2} A_t^2 - \sum_{n,m} \sum_s g^{2n+2m+s} \sum_L C(n,m,s,L) N^{-n-m-s+L} \left( {A_t \over N} \right)^n \left( {A_{t+1} \over N} \right)^m \right] ,  \nonumber \\
S_1 & = & -g \sum_t \left( N A_t + {1 \over 3N} A_t^3 \right).
\end{eqnarray}
Every factor of the form $A_t^n$ in Eq.(\ref{eq:effective}) has no tensor index left after the contraction.
Let us focus on the second term in the square bracket of $S_0$.
It is the sum over all products of two surface fields for any possible combination of square numbers $n$ and $m$ at time slices $t$ and $t+1$, respectively, with some weight.
Here, $n$, $m$, and $s$ are the numbers of up-pyramids, down-pyramids, and tetrahedra, respectively, while $L$ is the total number of the inner time-like links, along which the edges of simplices are fixed in a 3D triangulated unit layer.
The power index of $g$ expresses the volume of the layer, or total tetrahedron number, where one pyramid is counted as two tetrahedra.
Then, the coefficient $C(n,m,s,L)$ counts the number of different configurations for the discretization of the 3D space-time shell under any ``quadrangulation'' of 2D surfaces, with $n$, $m$, $s$, and $L$ fixed.
This is beyond the knowledge of the combinatorics, unlike the binomial coefficients in the case of the 2D model.
 
The power index of $N$ is found to be $-n-m-s+L=2-2h$, the Euler characteristic of the genus number $h$ for the closed surface propagating in this one-step time \cite{AJL1}.
We will confirm this relation in the appendix.
On the analogy of the 2D CDT model, the definition of the ``one-step propagator" is
\begin{eqnarray}
\label{eq:onestep}
 \sum_s g^{2n+2m+s} \sum_L C(n,m,s,L)\delta_{-n-m-s+L,2-2h} & \equiv & G^{(h)}(n,m) \hspace{10mm} \left( G^{(0)}(n,m)\equiv G(n,m) \right) \nonumber \\
 & \equiv & \sum_{\alpha} \sum_{\beta} G^{(h)}(n,m;\alpha , \beta ),
\end{eqnarray}
where the indices $\alpha$ and $\beta$ are assigned to each configuration of quadrangulation on the surfaces before and after the propagation, respectively.
A configuration with $n$ squares corresponds to the pairing of tensor indices in the contraction of the invariant $A_t^n$, and hence the field is distinguished with the additional index $\alpha$ as $\phi _t^{(h)} (n;\alpha )$ in the ``quadrangulated expression''.
When we use $\phi _t^{(h)} (n)$, i.e., ``area expression'' without the configuration index, it expresses any element of the set $\{ \phi _t^{(h)} (n;\alpha) | \alpha \in {\rm all~configurations} \}$ with its area $n$ fixed.
While $G^{(h)}(n,m;\alpha ,\beta )$ counts the number of configurations of the simplices after fixing the quadrangulation, $\alpha$ and $\beta$, of discretized surfaces, $G^{(h)}(n,m)$ sums up those for all the methods of quadrangulation with only the areas of surfaces $n$ and $m$ fixed.
The $S_0$ part of the effective action is rewritten in the quadrangulated expression as 
\begin{eqnarray}
\label{eq:effectiveactionS0}
S_{\rm 0} & = & \sum_t \left[  {1 \over 2}(A_t)_{abcd}(A_t)_{dcba}   - N^2 \sum_h \sum_{n,m} \sum_{\alpha , \beta} \phi _t^{(h)}(n;\alpha) G^{(h)}(n,m;\alpha,\beta) \phi _{t+1}^{(h)}(m;\beta) \right].
\end{eqnarray}
This expression is appropriate to realize the composition rule, since any propagator in the exact CDT from $\phi _0^{(h)} (n_0)$ to $\phi _t^{(h)} (n_t)$ in the interval of discrete time $t$ is obtained as
\begin{eqnarray}
\label{eq:propagator}
\langle \phi _0^{(h)} (n_0) \phi _t^{(h)} (n_t)\rangle _0 & = & \sum_{\alpha_0 , \alpha_t} \langle \phi _0^{(h)}(n_0;\alpha_0 ) \phi _t^{(h)}(n_t;\alpha_t ) \rangle _0 \nonumber \\
 & \equiv & \sum_{\alpha_0 , \alpha_t} {1 \over Z_0} \int {\cal D} A \phi _0^{(h)} (n_0;\alpha_0) \phi _t^{(h)} (n_t;\alpha_t) e^{-S_0} \nonumber \\
 & = & N^{(2-2h)t} \sum_{n_1,\cdot \cdot \cdot ,n_{t-1}=1}^{\infty} \sum_{\alpha_0, \alpha_1, \cdot \cdot \cdot ,\alpha_{t-1}, \alpha_t} \prod_{i=0}^{t-1} G^{(h)}(n_i, n_{i+1} ; \alpha_i , \alpha_{i+1} ),
\end{eqnarray}
where $Z_0$ is the partition function with only the $S_0$ part of the effective action.
The additional part of the action, $S_1$, causes the fluctuation in the propagation, with annihilation and creation of a square on the constructive surface field.
We will see that this relates to the surface propagation in one time slice with decreasing and increasing its size as a quantum effect, which distinctly violates the time-foliation structure.
Although it seems to disturb the construction of the CDT model, we will find it to play rather an important role in taking the continuum limit in the following sections.
Because our interest is the power index of the scaling parameter, or the scaling dimension, for each interaction term, we need to distinguish only the square number but not the configuration of the quadrangulation.
Hereafter, we describe fields and one-step propagators in the area expression.
Accordingly, the effective action is in the same form as the 2D model:
\begin{eqnarray}
\label{eq:effectiveaction}
S_{\rm eff} & = & N^2 \sum_t \left[ -g \left({A_t \over N}\right) + {1 \over 2} \left({A_t \over N}\right)^2 - {g \over 3} \left({A_t \over N}\right)^3 - \sum_h \sum_{n,m} \phi _t^{(h)}(n) G^{(h)}(n,m) \phi _{t+1}^{(h)}(m) \right] ,
\end{eqnarray}
where the contractions omitted in the first three terms are the same as the first three terms of Eq.(\ref{eq:action}).

\section{Stochastic quantization}
We apply the stochastic quantization method to this model to extend CDT to the GCDT model with a quantum effect containing the IK-type interaction, which also exists as well as the splitting interaction in the 2D model.
The Langevin equation of the fundamental tensor $A_t$ is
\begin{eqnarray}
\label{eq:langevin}
\Delta (A_t)_{abcd} & = & - {{\partial S_{\rm eff}} \over {\partial (A_t)_{dcba}}} \Delta \tau + (\Delta \xi _t)_{abcd}, \nonumber \\
& = & \Delta \tau \left[ Ng \delta_{ad} \delta_{bc} -(A_t)_{abcd} +{g \over N} (A_t)_{abfe} (A_t)_{efcd} \right. \nonumber \\
& & +\left. \sum_h \sum_{n,m}  n N^{1-h} \left( {A_t \over N} \right)^{n-1}_{abcd} \left\{ G^{(h)}(n,m)\phi_{t+1}^{(h)}(m) + G^{(h)}(m,n)\phi_{t-1}^{(h)}(m) \right\} \right] \nonumber \\
& & +(\Delta \xi_t)_{abcd}.
\end{eqnarray}
The last term is the white noise, which is normalized on account of the cyclic permutation of the correlation:
\begin{eqnarray}
\label{eq:correlation}
\langle (\Delta \xi _t)_{abcd} (\Delta \xi _{t'})_{d'c'b'a'} \rangle _{\xi}  & = & {1 \over 2} \Delta \tau \delta _{tt'}  (\delta _{aa'} \delta _{bb'} \delta_{cc'} \delta_{dd'} 
+\delta _{ba'} \delta _{cb'} \delta_{dc'} \delta_{ad'} \nonumber \\
&&~~~~~~~~~~~~+\delta _{ca'} \delta _{db'} \delta_{ac'} \delta_{bd'} +\delta _{da'} \delta _{ab'} \delta_{bc'} \delta_{cd'}  ).
\end{eqnarray}
The constructive surface field evolves in the one-step stochastic time $\Delta \tau$ as
\begin{eqnarray}
\label{eq:langevinphi}
\Delta \phi _t^{(h)} (n) &=& \Delta \tau n \left[ g \phi _t^{(h)} (n-1) - \phi _t^{(h)} (n) + g \phi _t^{(h)} (n+1)  + {1 \over N}(n-1) \phi _t^{(h+1)} ( n-2 ) \right. \nonumber \\
 & & \left. + \sum_{h'=0}^{\infty} \sum_{k=1}^{\infty} \phi _t^{(h+h')} (n+k-2) \sum_{m=0}^{\infty} \left\{ k G^{(h')}(k,m) \phi _{t+1}^{(h')} (m) + k G^{(h')}(m,k) \phi _{t-1}^{(h')} (m) \right\} \right] \nonumber \\
 & & + \Delta \zeta _t^{(h)} (n).
\end{eqnarray}
The last line is the constructive noise term defined by
\begin{eqnarray}
\label{eq:constructive}
\Delta \zeta _t^{(h)} (n) \equiv {n \over N^{h+1}} \left( {A_t \over N} \right)^{n-1}_{dcba} (\Delta \xi_t )_{abcd}
\end{eqnarray}
which satisfies the following correlation:
\begin{eqnarray}
\label{eq:correlationzeta}
\langle \Delta \zeta _t^{(h)} (n)  \Delta \zeta _{t'}^{(h')} (m) \rangle _{\xi}  =  2\Delta \tau \delta _{tt'} {1 \over N^2} nm \phi _t^{(h+h')} (n+m-2).
\end{eqnarray}
The second line of Eq.(\ref{eq:langevinphi}) expresses the creation of a baby closed surface with any square number $m$ at the neighboring times $t \pm 1$.
This creation happens simultaneously with the expansion of the original surface at $t$ with any square number $k-2$, weighted by the one-step propagator relating the $m$ and $k$ surfaces.
This is the IK-type interaction.

We define the FP Hamiltonian by the generator of the stochastic time evolution of any function of the fields $O(\phi )$,
$\langle \Delta O(\phi ) \rangle _\xi  \equiv  - \Delta \tau \langle H_{\rm FP} O(\phi ) \rangle _{\xi} $.
The lowest order in the expansion of the l.h.s. with respect to $\Delta \tau$ leads to the following formula:
\begin{eqnarray}
\label{eq:FPH}
H_{\rm FP} & = &  -{1 \over N^2} \sum_t \sum_{h=0}^{\infty} \sum_{n=1}^{\infty} n L_t^{(h)} (n-2) \pi _t^{(h)} (n) ,
\end{eqnarray}
where
\begin{eqnarray}
\label{eq:generator}
L_t^{(h)} (n) & \equiv & -N^2 \left[ g \phi _t^{(h)} (n+1) - \phi _t^{(h)} (n+2) + g \phi _t^{(h)} (n+3) + \sum_{h'=0}^{\infty} \sum_{k=1}^{\infty} \phi_t^{(h+h')} (n+k) \tilde{\phi}_t^{(h')} (k) \right. \nonumber \\
& & \left. +{1 \over N}(n+1) \phi _t^{(h+1)} (n) + {1 \over N^2} \sum_{h'=0}^{\infty} \sum_{k=1}^{\infty} k \phi_t^{(h+h')}(n+k) \pi_t^{(h')} (k) \right] .
\end{eqnarray}
In the above equation, $\pi_t^{(h)} (n) \equiv {\partial \over \partial \phi_t^{(h)} (n)}$ is the annihilation operator conjugated to the surface field creation operator $\phi_t^{(h)} (n)$, and hence they satisfy the canonical commutation relation, 
$[\pi_t^{(h)} (n) , \phi_{t'}^{(h')} (m)] = \delta _{tt'} \delta _{hh'} \delta_{nm}$. 
In addition, we have adopted the abbreviated field form for the set of fields created at the neighboring times in the IK-type interaction:
\begin{eqnarray}
\label{eq:tilde}
\tilde{\phi}_t^{(h)} (k) \equiv \sum_{m=0}^{\infty} \left\{ k G^{(h)}(k,m) \phi _{t+1}^{(h)} (m) + k G^{(h)}(m,k) \phi _{t-1}^{(h)} (m) \right\}.
\end{eqnarray}
As we expect, the generator $L_t^{(h)} (n)$ certainly satisfies the Virasoro-type algebra:
\begin{eqnarray}
\label{eq:virasoro}
\left[ L_t^{(h)} (n) , L_{t'}^{(h')} (m) \right] = (n-m) \delta_{t t'} L_t^{(h+h')} (n+m) .
\end{eqnarray}

Though the formulation of this model is similar to the 2D GCDT matrix model, the most distinct feature in the 3D model is the nonexistence of the splitting interaction, dividing one closed surface into two while conserving the total area.
In exchange, there exists the interaction that adds one handle on the surface, e.g., the deformation from a sphere to a torus or from a torus to a closed surface with two handles and so on.
This is caused by joining two squares at a distance from each other on the same surface, as in the splitting interaction of the 2D model.
Although the latter does not change the topology of the world sheet in 2D space-time, this handle-adding interaction in the 3D model brings on a topological change in the space-time geometry.
In the same way as with the 2D GCDT model, we permit the IK-type interaction, while not conceding the merging interaction.
\section{Continuum limit}
Now we will take the double-scaling limit to obtain the continuum model of GCDT from the above discrete model, which contains some uninvited interactions.
The length of all links, both space-like and time-like, is the minimal length $a$.
According to CDT, the area of the closed surface scales as $A \equiv a^2 n $ and the finite time or geodesic distance scales as $T \equiv at$.
Then the cosmological constant $\Lambda$ is related to the coupling constant $g$ as $g \equiv {1 \over 2} e^{-\Lambda a^3}$.
Here, we will introduce two scaling dimensions $D$ and $D_N$.
The latter concerns the scaling of $1/N^2$ to approach a coupling constant $G_{\rm st}$, whose subscript refers to the corresponding ``string coupling constant'' in the 2D string model:
\begin{eqnarray}
\label{eq:coupling}
G_{\rm st} \equiv a^{D_N} {1 \over N^2}.
\end{eqnarray}
The scaling indices of the other variables are expressed with linear functions of $D$.
The creation operators of the continuum spherical field (and the continuum closed surface field with $h$ handles) with area $A$ at time $T$ are defined by
\begin{eqnarray}
\label{eq:creation}
\Phi (A,T) \equiv a^{-{1 \over 2}D} \phi _t (n)~~~~\left({\rm and}~~~ \Phi ^{(h)} (A,T) \equiv a^{-{1 \over 2}D+{h \over 2}D_N} \phi _t^{(h)} (n) \right),
\end{eqnarray}
respectively.
We have assumed that the addition of one handle relates to the multiplication of $1/N$ in the discrete surface variable following Eq.(\ref{eq:invariants}), with Eq.(\ref{eq:coupling}). 
On the other hand, the annihilation operators of the corresponding fields are
\begin{eqnarray}
\label{eq:creation}
\Pi (A,T) \equiv a^{{1 \over 2}D-3} \pi _t (n)~~~~\left({\rm and}~~~ \Pi ^{(h)} (A,T) \equiv a^{{1 \over 2}D-{h \over 2}D_N -3} \pi _t^{(h)} (n) \right),
\end{eqnarray}
since these continuum fields have to satisfy the commutation relations
\begin{eqnarray}
\label{eq:canonical}
[\Pi ^{(h)} (A,T), \Phi ^{(h')} (A',T') ] =\delta _{hh'} \delta (A-A') \delta (T-T') .
\end{eqnarray}
Although the behavior of the one-step propagator $G^{(h)}(m,n)$ is analytically incomprehensible for this 3D model, we assume that the continuum counterpart of Eq.(\ref{eq:tilde}) effectively scales as a field; this assumption is exact for the 2D model:
\begin{eqnarray}
\label{eq:tilde2}
\tilde{\Phi} ^{(h)} (A,T) \equiv a^{-{1 \over 2}D+{h \over 2}D_N} \tilde{\phi} _t^{(h)} (n) .
\end{eqnarray}
The continuum limit of the FP Hamiltonian ${\cal H}_{\rm FP}$ and the infinitesimal stochastic time $d \tau$ is defined through the relation 
${\cal H}_{\rm FP} d \tau = H_{\rm FP} \Delta \tau $.

Because the sole candidate for the quantum effect is the IK-type interaction, we fix the scaling of $\Delta \tau$ as this interaction survives in the double-scaling limit.
Consequently, the continuum stochastic time is deduced:
\begin{eqnarray}
\label{eq:stochastict}
d \tau \equiv a^{{1 \over 2}D-4} \Delta \tau .  
\end{eqnarray}
The requirement of the stochastic time to be a continuum in the double-scaling limit leads to the condition
\begin{eqnarray}
\label{eq:condition1}
D>8. 
\end{eqnarray}
At this step, the continuum limit of the FP Hamiltonian is
\begin{eqnarray}
\label{eq:continuumFPH}
{\cal H}_{\rm FP}  & = & - \sum_{h=0}^{\infty} \int dT \int_0^{\infty} dA A \left[ - a^{-{1 \over 2}D+5} \Lambda \Phi ^{(h)} (A,T)  \right. \nonumber \\
 & & + \int _0^{\infty} dA_1 \Phi ^{(h)}(A+A_1,T) \tilde{\Phi} (A_1,T)  \nonumber \\
 & & + \sum_{h'=1}^{\infty} a^{-h' D_N}\int _0^{\infty} dA_1 \Phi ^{(h+h')}(A+A_1,T) \tilde{\Phi}^{(h')} (A_1,T)   \nonumber \\
 & & + a^{-{1 \over 2}D-{D_N}} \sqrt{G_{\rm st}} A \Phi ^{(h+1)} (A, T)  \nonumber \\
 & & \left.  + a^{-D - D_N +1} G_{\rm st} \sum_{h'=0}^{\infty} \int _0^{\infty} dA_1 A_1 \Phi ^{(h+h')} (A+A_1, T) \Pi ^{(h')} (A_1, T) \right]  \Pi ^{(h)}(A , T) . 
\end{eqnarray}
Let us estimate the scaling limit of each term under the condition that the second term, the IK-type interaction, scales to $a^0$ as mentioned above.
The second and third terms express a series of IK-type interactions.
The former expands a closed surface by the area $A_1$ without changing genus, simultaneously creating a sphere with the same area $A_1$.
This is interpreted as an interaction with the spherical domain wall with the area $A_1$.
On the other hand, the latter is the interactions with a domain wall whose shape is a closed surface with $h'(>0)$ handles.
Again, additional $h'$ handles on a surface field may change the scaling index naively by $-{1 \over 2} h' D_N$.
Hence the third line scales out relative to the second line under the condition of $ D_N<0$, which is rational for finite $G_{\rm st}$ from Eq.(\ref{eq:coupling}).

The first line means the propagation of the surface field in one time domain, which must not exist for the time-foliation structure in the GCDT model.
In order that this term scales out, we need
\begin{eqnarray}
\label{eq:condition2}
D<10. 
\end{eqnarray}
Some comments should be made about this scaling.
Thanks to the coexistence of the terms diminishing and increasing a square in the discrete level, or the first and third terms in Eq.(\ref{eq:langevin}), the scaling power index is enhanced three orders higher than that of each original term.
We could not make the propagation in a time domain forbidden, consistent with the former condition of Eq.(\ref{eq:condition1}), without this enhancement.
While in the 2D model the kinetic term, or the second derivative term, and the cosmological term coexist in the same order, in the 3D model the power of the second derivative term becomes higher than the cosmological term. 

The fourth line is the handle-adding interaction, which is rooted in the correlation of white noise square terms, as well as the splitting interaction permitted in GCDT in the 2D model. 
Meanwhile, in the 3D model, this interaction possesses the same problem as the light-cone degeneracy because it is the merging of two different points on the 2D space, or, substantially, the merging interaction.
Accordingly, we expect that it scales out with the condition
\begin{eqnarray}
\label{eq:condition3}
D_N <-{1 \over 2}D.
\end{eqnarray}

\begin{figure}[t]
\begin{center}
\includegraphics [width=80mm, height=65mm]{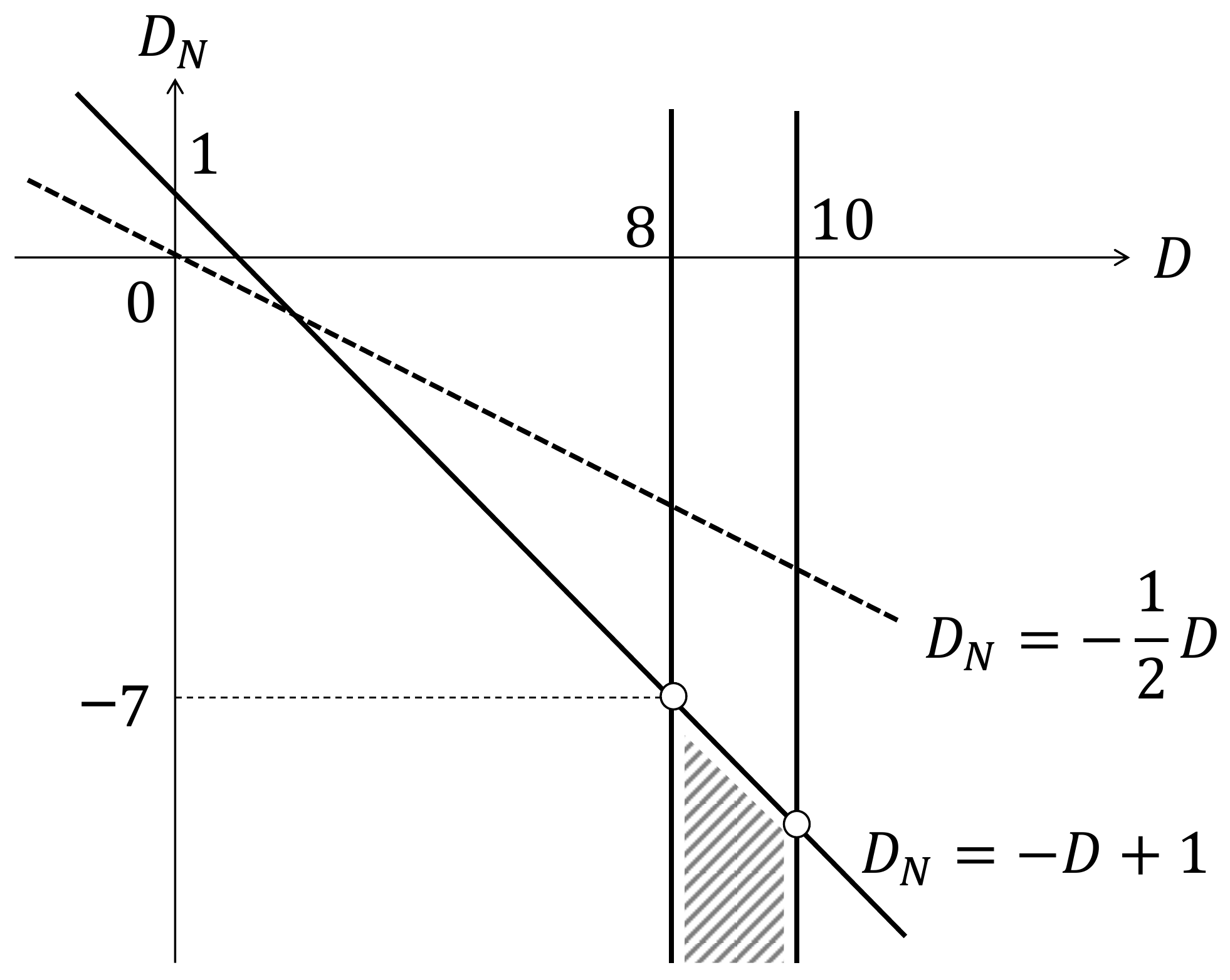}
\caption{In the shaded area in the $D_N$-$D$ space of the scaling dimensions, we obtain GCDT with the IK-type interaction. Whole the area is under the condition of $D_N < -{1 \over 2}D$. This means that, in the realization of our model, the handle-adding interaction is always forbidden.}
\label{fig:area}
\end{center}
\end{figure}

The last line is the merging interaction connecting two closed surfaces to make a large one with the totals of area and genera preserved.
The requirement of the merging interaction to be forbidden leads to the restriction
\begin{eqnarray}
\label{eq:condition4}
D_N<-D+1. 
\end{eqnarray}
Under the condition that all of Eqs.(\ref{eq:condition1}), (\ref{eq:condition2}), and (\ref{eq:condition4}) are satisfied, Eq.(\ref{eq:condition3}) is always fulfilled; hence we never have the handle-adding interaction in the continuum limit. 
In Fig.\ref{fig:area}, we find the appropriate domain in the $D_N$-$D$ space realizing the GCDT model with the IK-type interaction as the sole quantum effect.

\section{Discussion}
We have constructed a 3D GCDT model with the IK-type interaction by applying the stochastic quantization method to the tensor version of the loop gas model.
Our interest is focused on the realizability of the GCDT field theory in the double-scaling limit.
It is a 3D generalization of the 2D GCDT matrix model, as well as an extension of the 3D exact CDT model to include the quantum effect.
In this GCDT model, the stochastic time is not the geodesic distance, but the step of the quantum process.
One of the distinct features in the 3D model is the nonexistence of the splitting interaction, whose origin in the 2D model is the white noise correlation.
This fact is reasonable since the correlation of the white noise terms concerns only two distant points on a surface, while the splitting of the sphere could happen along a ring narrowing one part, which shrinks to a point.
The only remaining possible quantum correction creating baby universes is the IK-type interaction.
In the similar manner to the 2D matrix model, the scaling limit realizing the GCDT model with the IK-type interaction naturally excludes the handle-adding interaction.

This model is based on an assumption concerning the last terms of the effective action (\ref{eq:effectiveaction}).
The one-step propagators weighting the product of two closed surface fields in the neighboring times play an important role from the two viewpoints.
Firstly, by the composition of these terms through Wick's theorem, we obtain any propagator in the exact CDT for some set of initial and final closed surfaces.
In the quadrangulated expression, with the fields $\phi_t^{(h)}(n;\alpha )$ and the one-step propagators $G^{(h)}(n,m;\alpha ,\beta )$, we realize the composition rule of the time-foliation structure.
Secondly, on the other hand, they become the domain walls of the IK-type interaction in the stochastic process.
A common feature with the 2D matrix model appears in the area expression, $\phi_t^{(h)}(n)$ and $G^{(h)}(n,m)$, with which our investigation proceeds. 
Since the combinatorics about the 3D one-step propagator is beyond our comprehension, we are not able to read the scaling index from the formula.
Hence, we make the assumption in Eq.(\ref{eq:tilde2}), that the $\tilde{\phi}_t^{(h)}(k)$ defined by Eq.(\ref{eq:tilde}) possesses the same scaling property as the ordinary field $\phi _t^{(h)}(k)$, with the analogy of the 2D model.
Then, the scaling of the one-step propagator is $\tilde{G} (A , A') \equiv a^{-4} G(k,m)$, according to
\begin{eqnarray}
\label{eq:tilde3}
\tilde{\phi}_t (k) & = & \sum _m k \left\{ G(k,m) \phi _{t+1} (m) + G(m,k) \phi _{t-1} (m) \right\} \nonumber \\
& \equiv & a^{{1 \over 2}D} A \int dA' \left\{ \tilde{G} (A , A') \Phi (A',T+a) + \tilde{G} (A', A) \Phi (A',T-a) \right\} \equiv a^{{1 \over 2}D} \tilde{\Phi} (A,T).
\end{eqnarray}
This makes the IK-type interaction, the second line of Eq.(\ref{eq:continuumFPH}), scale regularly.
The IK-type interaction term in the continuum FP Hamiltonian is set in a similar form to the 2D model, with only the replacement of the argument by area $A$, instead of length $L$.

Even if the scaling of the one-step propagator differs from the above assumption, e.g., by $x$ or $\tilde{G} (A , A') = a^{-4+x} G(k,m)$, the IK-type interaction survives, with shifts of the conditions (\ref{eq:condition1}), (\ref{eq:condition2}), (\ref{eq:condition3}), and (\ref{eq:condition4}), as
\begin{eqnarray}
\label{eq:conditions}
D>8+2x, ~~~~D<10+2x, ~~~~D_N<-{1 \over 2}D+x,~~~~D_N<-D+1+x,  
\end{eqnarray} 
respectively.
The significant point is that the area satisfying above conditions remains, shifted in the space of scaling dimensions as well.

Finally, let us discuss the possibility of stable propagation for the closed surface field, or the 2D universe.
In the time-foliation structure of CDT, the universe propagates with its volume varying, or increasing and decreasing gradually.
In contrast, the IK-type interaction brings on a sudden expansion of the universe while simultaneously creating a baby universe. 
In the GCDT model, every baby universe eventually shrinks, to disappear without merging again with any other universe, including the mother universe.
Thus the universe tends to increase its volume by the IK-type interaction.
In the 2D GCDT model, we also have the ordinary splitting interaction: branching of a baby string from the mother string with their total length preserved.
Since the splitting interaction tends to decrease the mean length of the mother string, both effects of extension and shrinkage are expected to balance each other.
On the other hand, in the 3D GCDT model, we have only the volume-increasing IK-type interaction, without any interaction decreasing its volume instantaneously.
This quantum effect may stop the universe shrinking and disappearing rapidly.
Accordingly, it makes the mean volume of the propagating universe tend to be large.
This model has the possibility of shifting the phase transition line in the parameter space bounding the phase of stable propagation and that of the short-lifetime propagation.

\section*{Acknowledgements}
The suthor thanks the participants of the ``String Field Theory 2015'' workshop at Nara Women's University.
Discussions during the workshop were useful in completing this work.

\section*{Appendix: ~Topological relation}
We ascertain the topological relation of the handle number $h$ of any one-step discretized propagator for the closed surface, $-n-m-s+L =2-2h$, where $n$, $m$, and $s$ are the numbers of  up-pyramids, down-pyramids, and tetrahedra, respectively.
$L$ is the number of internal links, to which the edges of glued simplices commonly attach, or the time-like links connecting two closed surfaces of times $t$ and $t+1$.
\\

\begin{figure}[t]
\begin{center}
\includegraphics [width=80mm, height=25mm]{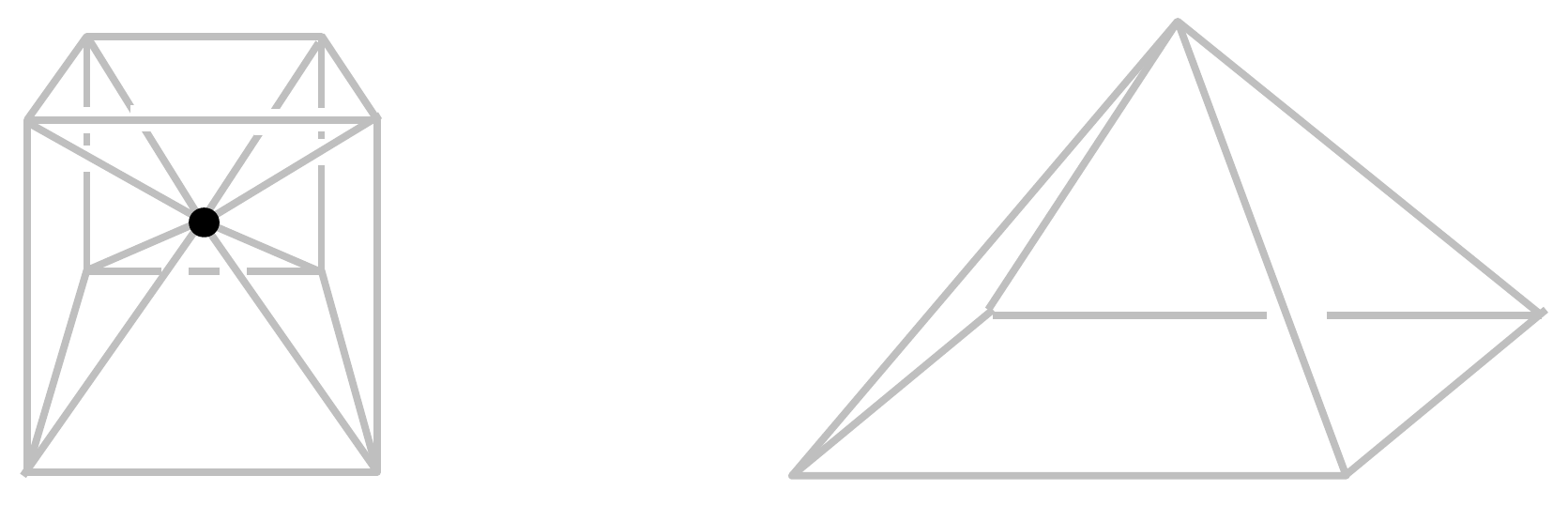}
\caption{A cube and a pyramid: A cube made of six pyramids is the simplest example of the annihilation of a sphere in the one-step propagation.}
\label{fig:sphere}
\end{center}
\end{figure}

We define an index $\chi$ as
\begin{eqnarray}
\label{eq:Euler}
\chi \equiv -n-m-s+L ,  
\end{eqnarray}
which we can easily count for some of the simplest one-step propagators.
For the first example, a cube made of six pyramids corresponds to the propagator from a closed surface of six squares to the point of a zero square, or the annihilation process (see Fig.\ref{fig:sphere}).
We can count six up-pyramids, no down-pyramid nor tetrahedron, and eight internal links, so $\chi = -6-0-0+8=2$ for this shell ($h=0$) in a broad sense. 

\begin{figure}[t]
\begin{center}
\includegraphics [width=25mm, height=25mm]{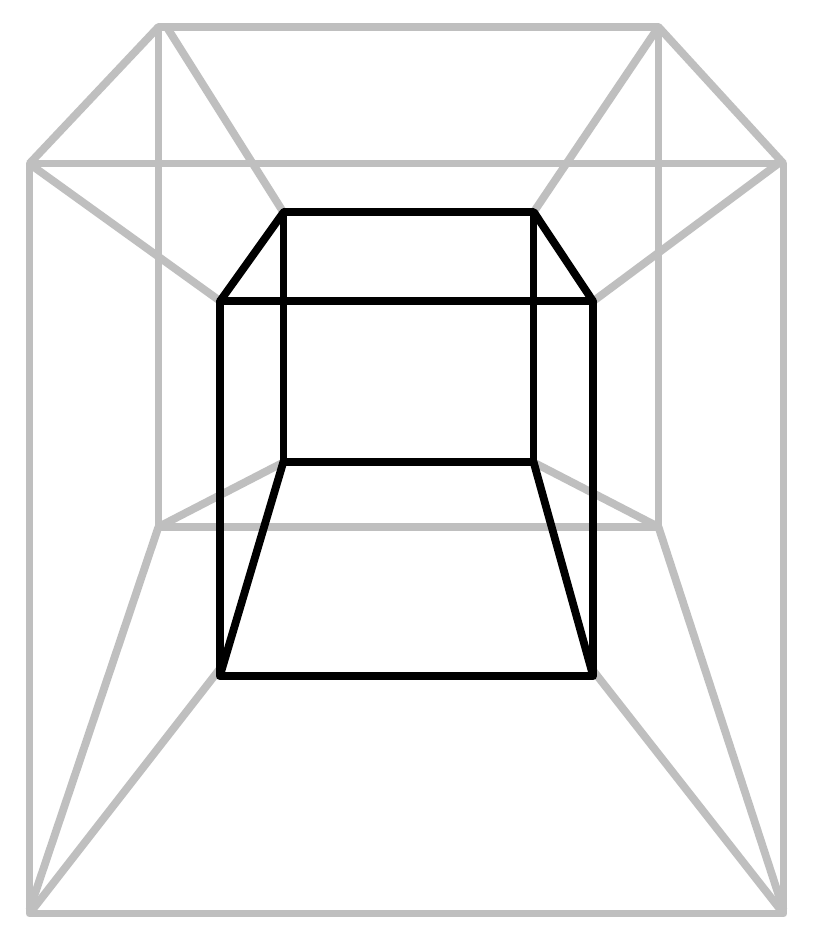}
\caption{A shell constructed by six footstools, each of which is formed by four up-pyramids, one down-pyramid, and four tetrahedra, as in Fig.\ref{fig:foliation}.}
\label{fig:shell}
\end{center}
\end{figure}

The second example is the simplest shell constructed by six footstools in Fig.\ref{fig:foliation} (see Fig.\ref{fig:shell}).
It is the one-step propagator from the closed surface with 24 squares to the one with six squares.
Each footstool of Fig.\ref{fig:foliation} contains four up-pyramids, one down-pyramid, four tetrahedra, and four internal links, at the step before attaching to the other ones.
After forming a cubic shell, the numbers of three types of simplex become just six times, while 32 new internal links occur in addition to the six times of the original link number.
Then we obtain $\chi = -24 -6 -24 +56 =2$ for this shell with no handle ($h=0$).\\

\begin{figure}[t]
\begin{center}
\includegraphics [width=70mm, height=50mm]{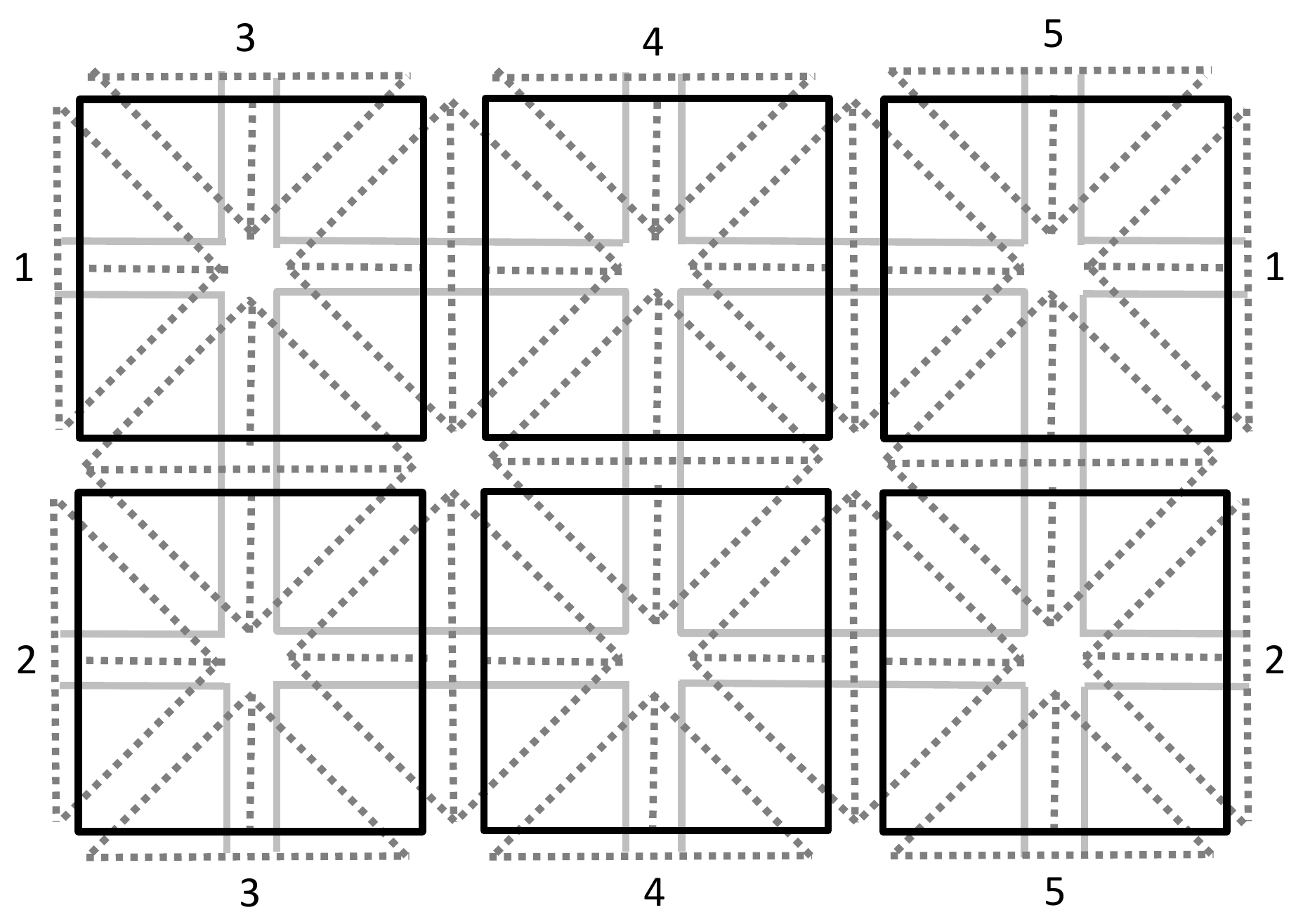}
\caption{A torus of two by three squares. The periodic boundary condition means two edges with the same number are identical. Oblique double broken lines correspond to the internal links.}
\label{fig:torus}
\end{center}
\end{figure}

The third example is a torus form of the one-step propagator, a pair of rectangles facing each other with the periodic boundary conditions for two directions (see Fig.\ref{fig:torus}).
For simplicity, we consider a rectangle made of two by three squares identifying five pairs of edges assigned with the same number.
The net numbers of six up-pyramids, six down-pyramids, twelve tetrahedra, and 24 internal links lead to $\chi = -6 -6 -12 +24 =0$ for this shell with one handle ($h=1$).
Of course, if we extend it to rectangles with $a$ by $b$ squares, the same manner of counting derives same value, $\chi = -ab -ab -2ab +4ab =0$. 

\begin{figure}[t]
\begin{center}
\includegraphics [width=70mm, height=35mm]{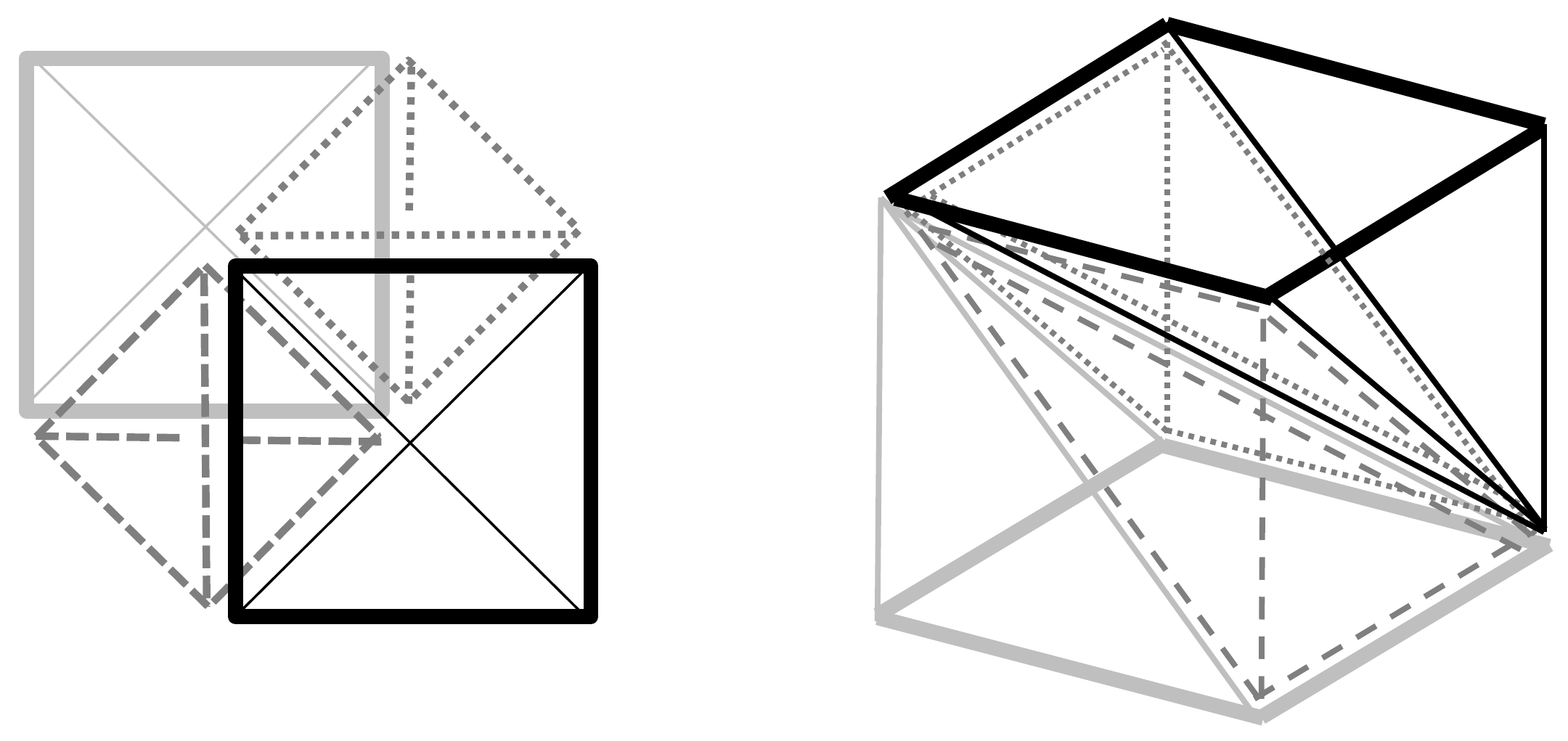}
\caption{The most fundamental hexahedron: The left figure is the picture seen from the upper direction. The right figure is the deformed expression corresponding to the cube containing one square in each of the $t$ and $t+1$ time slices. It contains two pyramids of solid lines, one black and the other gray, and two tetraredra shown by two kinds of broken lines.}
\label{fig:cube}
\end{center}
\end{figure}

Now we consider the fundamental cube, properly a hexahedron, constructed by one up-pyramid, one down-pyramid, and two tetrahedra, as in Fig.\ref{fig:cube}.
The above extension from the two by three torus to the $a$ by $b$ torus is made by adding $ab-6$ cubes.
Each additional cube creates on average four new internal links, which causes no variation of the index $\chi$, $\Delta \chi =-1 -1 -2 +4 =0$.
Until now, we have seen that Eq.(\ref{eq:Euler}) is satisfied for the above three types of the most basic shell, i.e., a shell whose inner surface shrinks to a point, a spherical shell, and torus shells.
\\

\begin{figure}[t]
\begin{center}
\includegraphics [width=70mm, height=30mm]{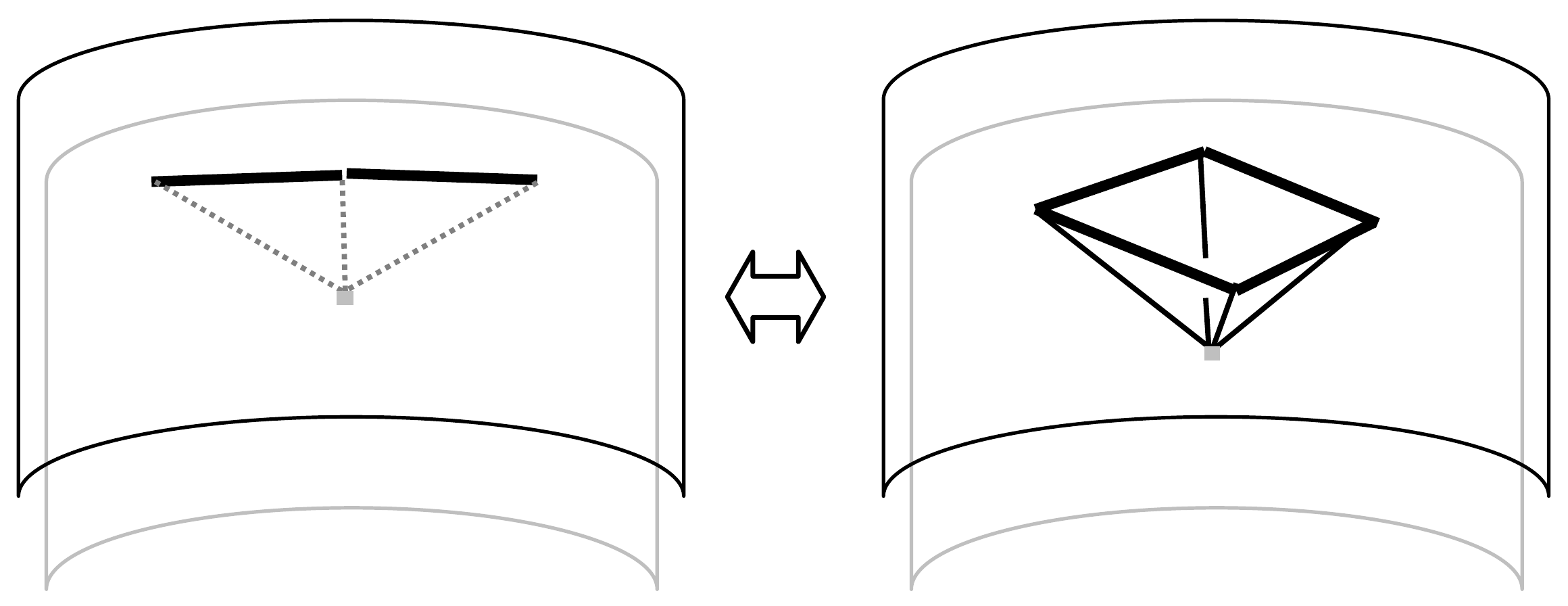}
\caption{Type I: The insertion of only one down-pyramid (up-pyramid) is possible at the cross section where the attached triangles align as $CC$ ($BB$). The opposite direction is the removal of one pyramid.}
\label{fig:quantum04}
\end{center}
\end{figure}

\begin{figure}[t]
\begin{center}
\includegraphics [width=70mm, height=30mm]{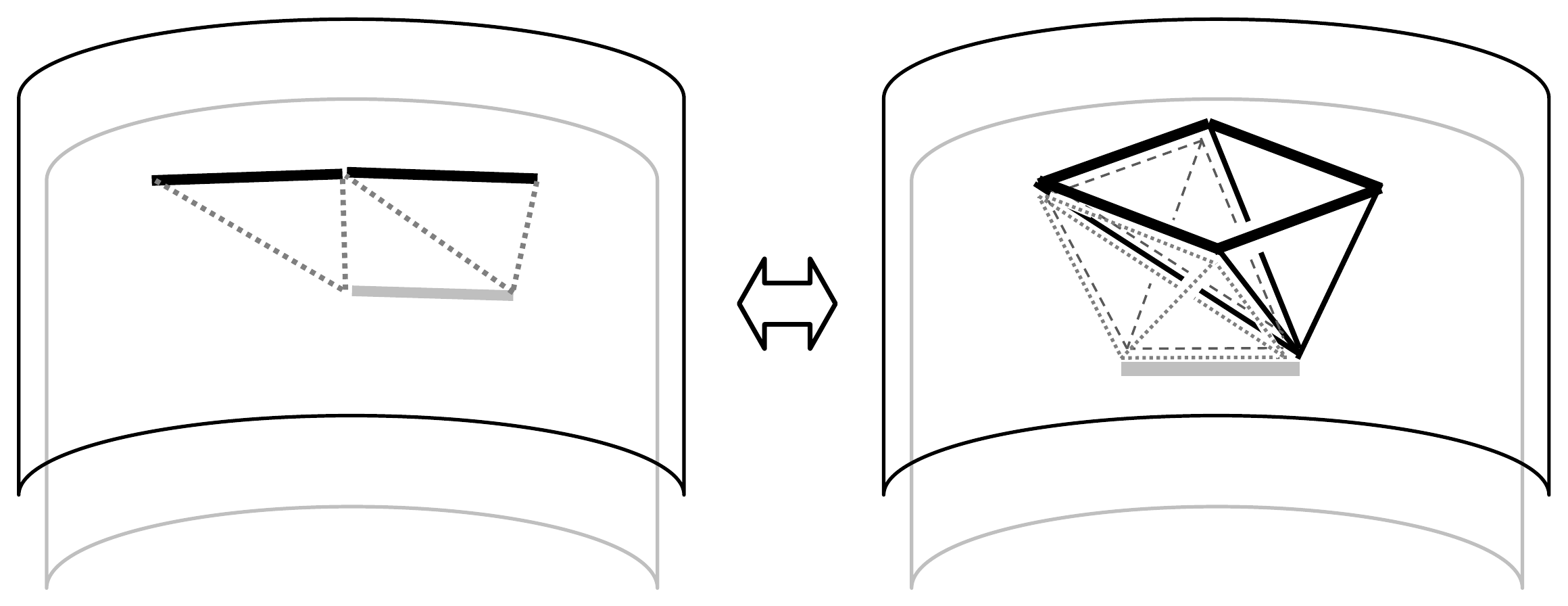}
\caption{Type II: The insertion of two tetrahedra in addition to one down-pyramid (up-pyramid) is possible at the cross section where the attached triangles align as $CBC$  ($BCB$). The opposite direction is the removal of the set of one pyramid and two tetrahedra.}
\label{fig:quantum14}
\end{center}
\end{figure}

\begin{figure}[t]
\begin{center}
\includegraphics [width=70mm, height=30mm]{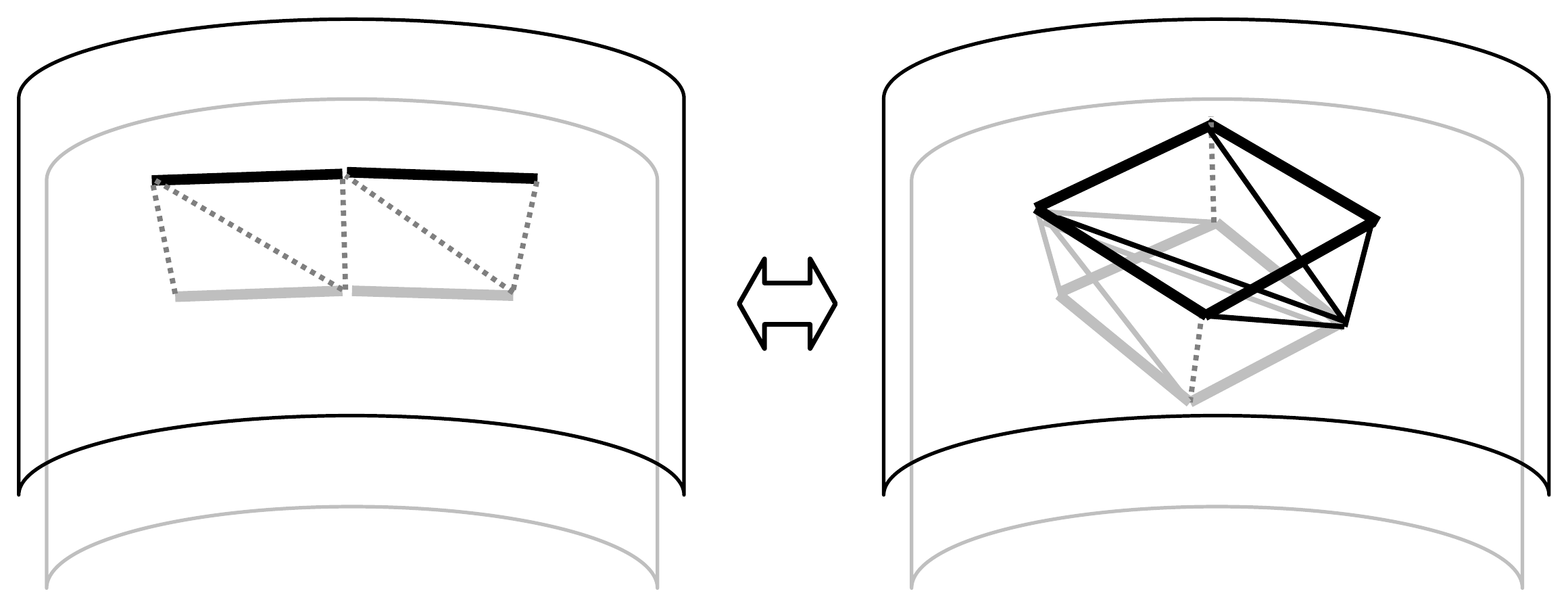}
\caption{Type III: The insertion of one cube of Fig.\ref{fig:cube}, or the set of one up-pyramid, one down-pyramid, and two tetrahedra, at the cross section where the attached triangles align as $BCBC$. The opposite direction is the removal of a cube.}
\label{fig:quantum44}
\end{center}
\end{figure}

Now let us consider the deformation of a discretized shell.
The fundamental deformation is the insertion of a pyramid, possibly with additional tetrahedra, at the attached triangles in a cross section.
When we insert one pyramid into the shell, we cut the discrete surface surgically along two consecutive links and open it there to make a loop of four links, occupied by the square of a pyramid.
Then, the detached triangles in the shell have to be attached to new triangles of the inserted simplices consistently.
We have the following three types of elementary insertion, with which index $\chi$ is invariant:
\begin{itemize}
\item Type I:~~The insertion of one pyramid only, or the propagator from a site to a square (and vice versa), at $CC$ (and $BB$) cross section (see Fig.\ref{fig:quantum04}).
\item Type II:~~The insertion of two tetrahedra in addition to one pyramid, or the propagator from a link to a square (and vice versa), at $BCB$ (and $CBC$) cross section (see Fig.\ref{fig:quantum14}).
\item Type III:~~The insertion of a cube of Fig.\ref{fig:cube}, or the propagator from a square to another square, at $BCBC$ cross section (see Fig.\ref{fig:quantum44}).
\end{itemize}
In each deformation we include the one in the opposite direction, or the removal of the set of simplices, as well.

Type I deformation is possible at a cross section where two triangles of the same type align, as $CC$ (or $BB$).
The insertion of one down-pyramid (or one up-pyramid) without changing the tetrahedron number is accompanied by the increase of one internal link, and consequently the variation of the index $\chi$ becomes $\Delta \chi =-1 -0 -0 +1 =0 ~(\Delta \chi = -0-1-0+1=0)$.
Type II deformation is possible at the cross section of three triangles where both types align in turn, as $CBC$ (or $BCB$).
The insertion of one down-pyramid (or one up-pyramid) in addition to two tetrahedra is related to the increase of three internal links; hence the value of $\chi$ is again kept, $\Delta \chi = -1-0-2+3=0 ~(\Delta \chi = -0-1-2+3=0)$.
Type III deformation is possible at the cross section of four triangles where both types align in turn, as $BCBC$ (or $CBCB$).
It is not exactly fundamental, because Type III is realized by the successive deformation of Type II and Type I at the cross section of $BCBC$.
Namely, after the Type II insertion for the $CBC$-part, Type I insertion follows at the $BB$-part; one is the remaining $B$ and the other is the up-triangle created inside the inserted part in Fig.\ref{fig:quantum14} by the preceding Type II.
Of course, this deformation does not change the index, $\Delta \chi = -1-1-2+4=0$, as mentioned above.

A discrete shell with any square numbers on both the surfaces may be realized through the combination of Type I and Type II deformation, as well as the insertion and exclusion of a series of cubes, from an arbitrary shell of the same topology.
For example, the footstool of Fig.\ref{fig:foliation} is inserted at the $BCBBCB$ part of the cross section, with twice Type II, twice Type I on the lower surface and once Type I on the upper surface in order, without changing the index $\chi$, $\Delta \chi = -4-1-4+9=0$.
While any discrete spherical shell is made through the appropriate deformation as above from the cubic shell of Fig.\ref{fig:shell}, it is possible to obtain any discrete shell with one handle by an appropriate method of deformation from the basic torus shell of Fig.\ref{fig:torus}.
We see that the index $\chi$ depends on the topology of the one-step propagator.\\

\begin{figure}[t]
\begin{center}
\includegraphics [width=70mm, height=30mm]{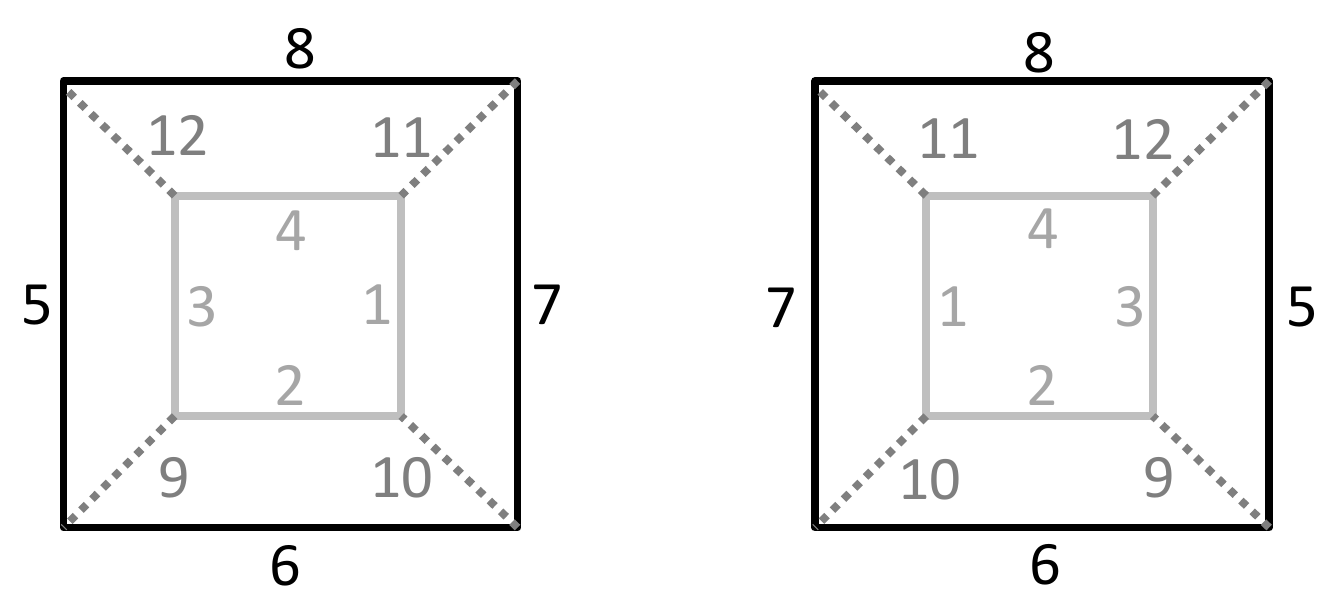}
\hspace{10mm}
\includegraphics [width=70mm, height=40mm]{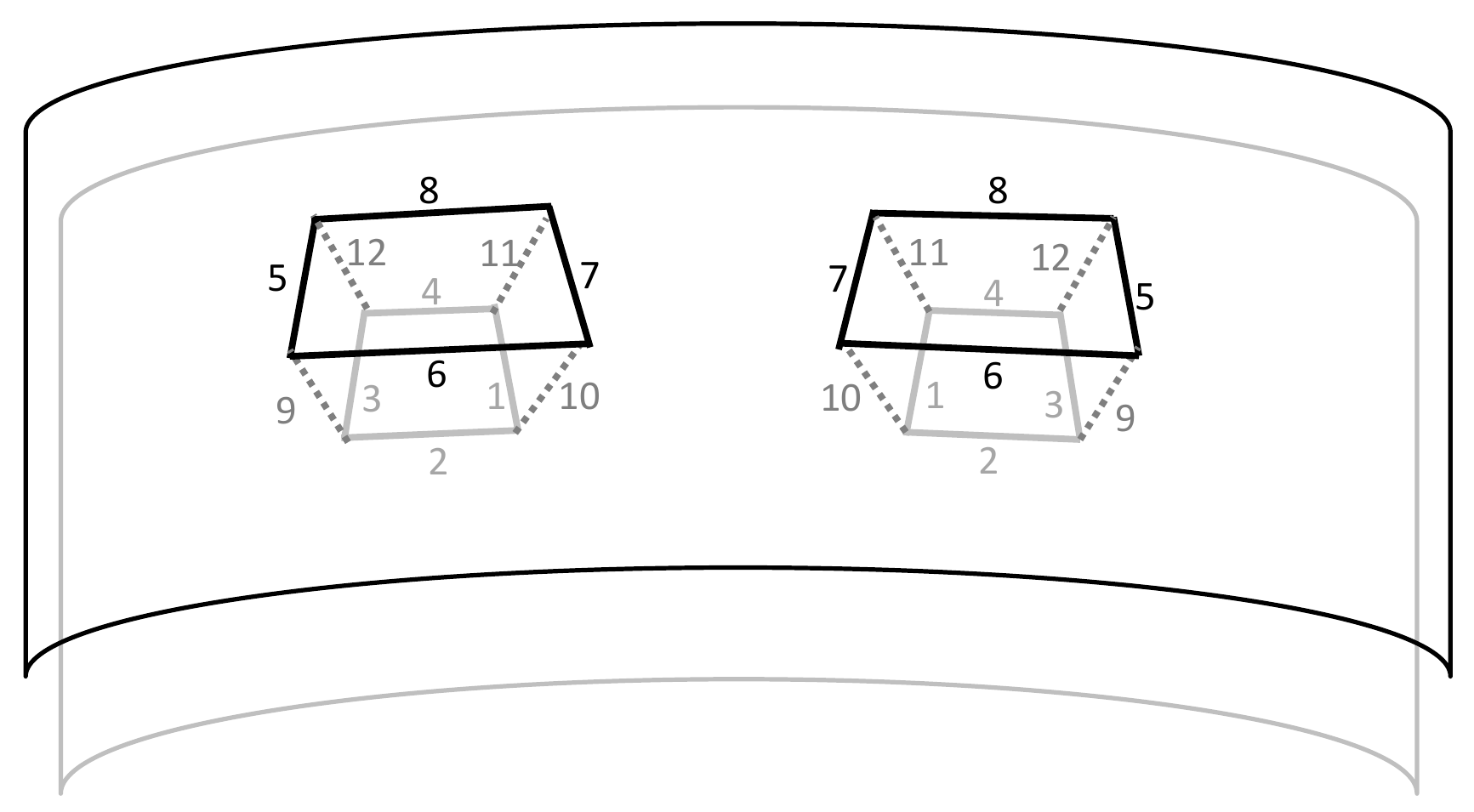}
\caption{Type IV: The process of adding one handle to the shell of the surface: The edges with the same number are attached. Two pairs of space-like squares with the edges 1234 on the surface of $t$ and with edges 5678 on $t+1$ are erased. Each time-like squares possesses one internal link, which is abbreviated in the above figures.}
\label{fig:handle}
\end{center}
\end{figure}

Finally, we add one more fundamental deformation:
\begin{itemize}
\item Type IV:~~The handle-adding deformation, or the pair annihilation of two cubes, with attaching time-like squares, or pairs of triangles, around them (see Fig.\ref{fig:handle}).
\end{itemize}
The handle-adding deformation, or the topology change, is caused by the white noise correlation at the level of the surface, which is now promoted to the level of the shell.
Consider two cubes like Fig.\ref{fig:cube} located at a distance from each other in the layer of a discrete closed surface, as in Fig.\ref{fig:handle}.
Through the identification of pairs of links with the same number, two cubes disappear and, simultaneously, four pairs of squares, 1$\cdot$10$\cdot$7$\cdot$11, 2$\cdot$9$\cdot$6$\cdot$10, 3$\cdot$12$\cdot$5$\cdot$9, and  4$\cdot$11$\cdot$8$\cdot$12, are attached to each other. 
In this deformation, two up-pyramids, two down-pyramids, four tetrahedra, and ten internal links disappear; hence the variation of the index $\chi$ is $\Delta \chi = -(-2)-(-2)-(-4)-10=-2$, which is accompanied by handle-number increase of one.\\

We summarize the discussion in this appendix:
\begin{enumerate}
\item For the basic discrete one-step propagator of a closed surface with no handle (including tadpoles), $\chi =2$.
\item For the basic discrete one-step propagator with the form of a torus, $\chi =0$.
\item The deformation of a discrete one-step propagator to any other one with the same handle number does not change the index $\chi$.
\item When we add one handle to a discrete one-step propagator, the value of $\chi$ changes by $\Delta \chi =-2$.
\end{enumerate}
Therefore, we see that, for any discrete one-step propagator with handle number $h$, the identity $\chi =2-2h$ is always satisfied.



\end{document}